# Web-page Prediction for Domain Specific Web-search using Boolean Bit Mask

Sukanta Sinha[1, 4], Rana Duttagupta[2], Debajyoti Mukhopadhyay[3, 4]

[1]Tata Consultancy Services, Victoria Park Building, Kolkata 700091, India
[2] Computer Science Department, Jadavpur University, Kolkata 700032, India
[3] Maharashtra Institute of Technology, Pune 411038, India
[4]WIDiCoReL, Green Tower C- 9/1, Golf Green, Kolkata 700095, India
E-mail: sukantasinha2003@gmail.com, rdattagupta@cse.jdvu.ac.in,
debajyoti.mukhopadhyay@gmail.com

**Abstract.** Search Engine is a Web-page retrieval tool. Nowadays Web searchers utilize their time using an efficient search engine. To improve the performance of the search engine, we are introducing a unique mechanism which will give Web searchers more prominent search results. In this paper, we are going to discuss a domain specific Web search prototype which will generate the predicted Web-page list for user given search string using Boolean bit mask.

## 1 Introduction

Search Engine is an information retrieving system of World Wide Web (WWW) [1]. The features of the search engine have become very complex. Web page prediction is an important feature of search engine, which produces search result more accurately for a user given search string.

In this paper, we will discuss the basic idea of domain specific search from Index Based Acyclic Graph (IBAG) using Boolean bit mask. We will also describe a design and development methodology for generation of bit pattern for all the Web-pages existing in IBAG and dynamic Web-page prediction list.

This paper discusses Web-page prediction in section 2. Section 3 tells about the existing model of Relevant Page Graph (RPaG) model and IBAG model. Section 4 depicts the proposed approach. Section 5 shows some performance analysis. Finally, section 6 concludes the paper.

**Definition 1.** Seed URL – It is a set of base URL from where the crawler starts to crawl down the Web pages from Internet.

**Definition 2.** Weight Table - This table contains two columns; first column denotes Ontology terms and second column denotes weight value of that Ontology term. Weight value must be within '0' and '1'.

**Definition 3.** Syntable - This table contains two columns; first column denotes Ontology terms and second column denotes synonym of that ontology term. For a particular ontology term, if more than one synonym exists then it should be kept using comma (,) separator.

## 2   Web-page Prediction

Web-page prediction implies predicting proper Web-page based on the given search string. The exponential proliferation of Web usage has dramatically increased the volume of Internet traffic and has caused serious performance degradation in terms of user latency and bandwidth on the Internet [2]. Web-page prediction [3] that involves personalizing the Web users' browsing experiences assist Web masters in the improvement of the website structure and helps Web users in navigating the site and accessing the information they need. Various attempts has been exploited to achieve Web-page access prediction by pre-processing Web server log files and analyzing Web users' navigational patterns. The most widely used approach for this purpose is Web usage mining that entails many techniques like Markov model, association rules, clustering, etc. [4]. In this paper, we are going to introduce a new method of domain specific Web-page prediction using Boolean bit mask. In our approach, we are mainly using bit wise exclusive-OR operation for finding predicted Web-page list [5].

## 3   Existing Models

In this section, we will describe two existing models; RPaG model, IBAG model.

**Definition 4.** Relevance Value – It is a numeric value for each Web-page; which is generated on the basis of the term Weight value, term Synonyms, Number of occurrence of Ontology terms are existing in that Web-page.

**Definition 5.** Relevance Limit – It is a predefined static relevance cut-off value to recognize whether a Web-page is domain specific or not.

**Definition 6.** Term Relevance Value – It is a numeric value for each Ontology Term; which is generated on the basis of the term Weight value, term Synonyms, Number of occurrence of that Ontology term in the considered Web-page.

**Definition 7.** Term Relevance Limit – It is a predefined static relevance cut-off value for each Ontology Term.

## 3.1 Relevance Page Graph Model

In this section, RPaG [6] is described along with the concept of its generation procedure. Every crawler needs some seed URLs to retrieve Web-pages from World Wide Web (WWW). All Ontologies, weight tables and syntables [7] are needed for retrieval of relevant Web-pages. RPaG is generated only considering relevant Web-pages. Each node in RPaG holds Web-page information. In RPaG, each node contains Page Identifier (P_ID), Unified Resource Locator (URL), four Parent Page Identifiers (PP_IDs), Ontology relevance value (ONT_1_REL_VAL, ONT_2_REL_VAL, ONT_3_REL_VAL), Ontology relevance flag (ONT_1_F, ONT_2_F and ONT_3_F) and Ontology terms relevance value (ONT_1_TERM_REL_VAL, ONT_2_TERM_REL_VAL and ONT_3_TERM_REL_VAL) fields information. "Ontology Relevance Value" contains calculated relevance value if these value grater than "Relevance Limit Value" of their respective domains. Otherwise, these fields contain "Zero (0)".

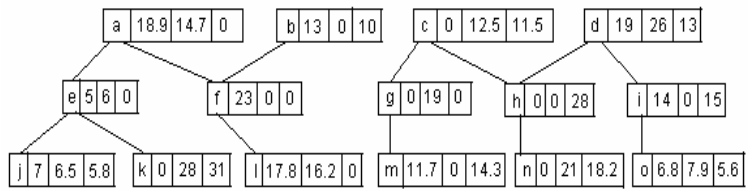

**Fig.1.** Arbitrary example of Relevance Page Graph

If page P supports 'Ontology 1'; i.e., relevance value grater than relevance limit; then 'Ontology 1' flag (ONT_1_F) must be 'Y'. Same way we also define 'Ontology 2' flag (ONT_2_F) 'Ontology 3' flag (ONT_3_F). 'Ontology 1' each term relevance value (ONT_1_TERM_REL_VAL) of Page P is generated according to the 'Ontology 1'. Similarly, 'Ontology 2' each term relevance value (ONT_2_TERM_REL_VAL) and 'Ontology 3' each term relevance value (ONT_3_TERM_REL_VAL) are generated according to the 'Ontology 2' and 'Ontology 3'. A sample RPaG is shown in Fig. 1. Each node in this figure of RPaG contains four fields; i.e., Web-page URL, ONT_1_REL_VAL, ONT_2_REL_VAL and ONT_3_REL_VAL.

## 3.2 IBAG Model

An acyclic graph is a graph having no graph cycles. A connected acyclic graph is known as a tree. IBAG [8] means an indexed tree. IBAG is typically generated from RPaG. In Fig. 2, a sample IBAG is shown. RPaG pages are related in some Ontologies and the IBAG generated from this specific RPaG is also related to the same Ontologies. Each node in the figure (refer Fig. 2) of IBAG contains Page Identifier (P_ID), Unified Resource Locator (URL), Parent Page Identifier (PP_ID), Mean Relevance value (MEAN_REL_VAL), Ontology 1 link (ONT_1_L), Ontology 2 link (ONT_2_L) and Ontology 3 link (ONT_3_L) fields. Along with those fields we also transfer ONT_1_TERM_REL_VAL, ONT_2_TERM_REL_VAL and

ONT_3_TERM_REL_VAL field information while generating IBAG from RPaG. Page Identifier (P_ID) is selected from RPaG page repository. Each URL has a unique P_ID and the same P_ID of the corresponding URL is mentioned into IBAG page repository.

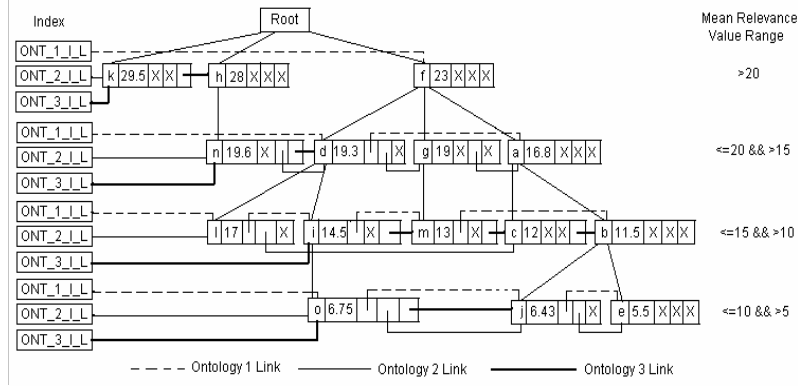

**Fig.2.** IBAG Model

Consider, one page supports 'Ontology 1' and 'Ontology 2'; then we calculate MEAN_REL_VAL as (ONT_1_REL_VAL + ONT_2_REL_VAL)/2. If one page supports 'Ontology 1', 'Ontology 2' and 'Ontology 3'; then we calculate MEAN_REL_VAL as (ONT_1_REL_VAL + ONT_2_REL_VAL + ONT_3_REL_VAL) / 3. 'Ontology 1 link' (ONT_1_L) points to the next 'Ontology 1' supported page. Similarly, 'Ontology 2 link' (ONT_2_L) points to the next 'Ontology 2' supported page. 'Ontology 3 link' (ONT_3_L) points to the next 'Ontology 3' supported page. In Fig. 2, we have shown only five fields; i.e., Web-page URL, MEAN_REL_VAL, ONT_1_L, ONT_2_L and ONT_3_L. In each level, all the Web-pages' "Mean Relevance Value" are kept in a sorted order and all the indexes which track that domain related pages are also stored.

**Definition 8.** Ontology – It is a set of domain related key information, which is kept in an organized way based on their importance.

## 4 Proposed Approach

In our approach, we have generated bit pattern for all Web-pages existing in IBAG and further have searched the Web-pages from IBAG for a given "Search String" to generate predicted Web-page list. Initially, we have generated RPaG which was constructed from typical crawling technique and then we have generated IBAG from that RPaG. After construction of IBAG we have generated bit pattern of all Web pages. Finally, a search string is given as input on the Graphical User Interface (GUI) along with other details; and as a result, corresponding list of predicted Web-page URLs is produced.

### 4.1 Bit Pattern Generation Algorithm

Bit pattern generation is a one time job. This job generates Web-page bit pattern for all Web-pages existing in IBAG. IBAG, Ontology terms relevance limit ($OT_{lmt}$) and number of Ontology term (t) for the taken Ontology are considered as input for this job. Each Ontology term holds a fixed position in the generated Web-page bit pattern. Those positions are absolutely predefined.

**Method 1.1:** Web-page Bit Pattern Generation
**genWebpageBitPatrn (Web-page, Ontology)**

```
1. generate dummy bit pattern with t number of 0's;
2. for each Ontology term perform 3-6;
3. fetch Ontology term limit (OT_lmt) for the selected
   Ontology Term;
4. Calculate Ontology term relevance value for the
   considered Web-page;
5. check Ontology term relevance value > OT_lmt then perform
   6 else goto 2;
6. change bit value 0 to 1 in dummy bit pattern for the
   selected Ontology term position;
7. store dummy bit pattern of selected Web-page for the
   corresponding Ontology;
```

**Algorithm 1:** IBAG Model Web-page Bit Pattern Generation
**genIBAGWebpageBitPatrn(IBAG Model Web-page, Ontologies)**

```
1. for each Web-page perform 2-3;
2. for each Ontology term perform 3;
3. call genWebpageBitPatrn(Web-page, Ontology);
```

For example, take one Web-page 'P' from IBAG. Now based on our Algorithm 1, first we have to generate a bit pattern of the Web-page 'P' which contains 't' number of bits, where 't' denotes number of ontology terms taken for the considered domain. Then check each ontology term relevance value with predefined ontology term relevance limit. Suppose, for Web-page 'P' 2nd and 5th ontology term exceeds the term relevance limit value. Hence the bit pattern of Web-page 'P' becomes like (0100100… t times).

### 4.2 Find Predicted Web-Page List

Predicted Web-page list is generated at runtime. This operation is performed for each search action. Initially we select the Web-pages from IBAG for the user given relevance range and selected Ontology. Then apply mask bit pattern to their respective Ontology to find whether the selected page belongs to predicted page list or not. Finally, predicted Web-page URLs are shown on the particular Web-page as the search result. Search string, relevance range, selected Ontology, number of search

result, IBAG, Web-page bit pattern and number of Ontology terms are considered as input for this procedure

**Method 2.1:** Mask Bit Pattern Generation
**getMaskBitPattern(Search String, Ontology)**

1. extract Ontology terms in search string;
2. create a Mask Bit Pattern by taking 1's for Ontology Terms present in search string and 0's for not present in search string and length must be t;
3. return maskBitPattern;

**Method 2.2:** get Web-page list from IBAG based on the user given relevance range
**getWebpagesFromIBAG(IBAG,relevanceRange,selectedOntology)**

1. generate Web-page list by traversing IBAG for the user given relevance range and selected Ontology;
2. return Web-pageList;

**Algorithm 2:** Find Predicted Web-Page List
**findPredictedWebpageList (IBAG, search string, relevance range, selected Ontology, Web-page bit pattern, number of search result)**
1. β := **getMaskBitPattern** (Search String, Ontology);
2. selectedWebpageList := **getWebpagesFromIBAG**(IBAG, relevanceRange, selectedOntology);
3. for each Web-page in selectedWebpageList perform 4-10
4. calculate µ = α ^ β;
5. for each Ontology term in search string perform 6-9
6. if (Ontology term position in µ = 0) then perform 7-9
7. add Web-page in predicted Web-page list;
8. predicted Web-page counter ++;
9. exit step-5 for loop;
10. if (predicted Web-page counter >= number of search result) then exit step-3 for loop;
11. display predicted Web-page list;

Where,
µ = Resulted Bit Pattern
α = Traverse Page Bit Pattern for the considered Ontology
β = Mask Bit Pattern

Now, for a given search string we have to find whether Web-page 'P' (refer algorithm1 example) needs to be included in the predicted Web-page List or not. Based on our Method 2.1, we have to create a mask bit pattern. Suppose, 2nd position ontology term exists in search string then the mask bit pattern looks like (0100000… t times). Again we assumed that the Web-page 'p' belongs to user given IBAG relevance range and supports user selected Ontology. Then as per our Algorithm 2, we perform XOR (^) operation between bit pattern of Web-page 'P' and mask bit pattern, i.e., (0100100… t times) ^ (0100000… t times) and the resultant bit pattern becomes (0000100… t times). Now we check 2nd position of the resultant bit pattern, if it is '0' then include the Web-page else discard Web-page 'P'. In our example, 2nd position of the resultant bit pattern showing zero (0), hence we include Web-page 'P' in predicted Web-page List.

## 5  Performance Analyses

Here we will explain our test setting and will also discuss some comparative study in our test result section.

### 5.1 Test Setting

In this section we have described weight table, Syntable and also explained our testing procedure. Weight table and syntable are used for calculating term relevance value. In Table 1 and Table 2 we have shown a sample weight table and Syntable for few Ontology terms.

Table 1. Weight Table

| | |
|---|---|
| cricket | 0.9 |
| wicket keeper | 0.8 |
| umpire | 0.4 |
| bat | 0.2 |
| match | 0.1 |

Table 2. Syntable

| | |
|---|---|
| Match | competition, contest |
| Stamp | stick, wicket |
| Ball | conglobate, conglomerate |
| Umpire | judge, moderator, referee |
| Catch | capture |

**Testing Procedure.** For experimental purpose, we have a set of search string, which we applied to both IBAG models; i.e., before bit masking and after bit masking, for our comparative study. First we have taken such an IBAG model which contains 1000 URLs. Now we applied all search strings to find search time taken and number of page retrieved by both models, which contains 1000 URLs in each model. Then we average search time for each model and plot the graph and also do the same for the number of pages. Same way we have taken 2000, 3000, 4000 and 5000 URLs to calculate search time and number of pages retrieved for plotting the graph. Finally from the graph we find the performance of our system.

### 5.2 Test Results

In this section we have generated some test results based on our test procedure and represented them by the graph plot. We have also verified accuracy of our search result after retrieval of predicted Web page list based on our given set of Search String. Accuracy measurement is determined based on some parameters like meaning of the Web page content, number of Ontology terms of that particular domain existing in the Web page content etc. Meaning of the Web page content is explained by seeing the content of the Web page and this is a manual process.

**Average Number of Predicted Web Page List for a Set of Search String.** In Fig. 3 we have shown average number of predicted Web pages retrieved from both IBAG models; i.e., before bit masking and after bit masking, for a given set of search string and various relevance rage values. From the figure we found, number of Web pages retrieved from "after bit masking in IBAG Model" is lesser than number of Web page retrieved from "before bit masking in IBAG Model."

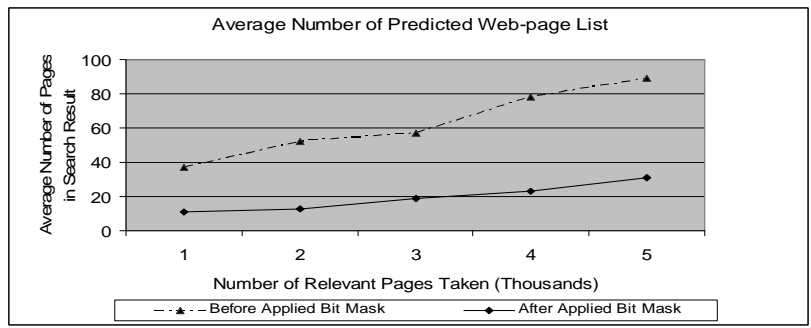

**Fig.3.** Comparison between Average Number of Web-pages Retrieved from before and After Bit Masking in IBAG Model

**Average Time Taken for a Set of Search String.** In Fig. 4 we have shown average time taken by both IBAG models; i.e., "before bit masking" and "after bit masking," for a given set of search string. From the figure we found that both IBAG models have taken near about same time but the accuracy of resultant predicted Web-pages list is better than before masking IBAG model.

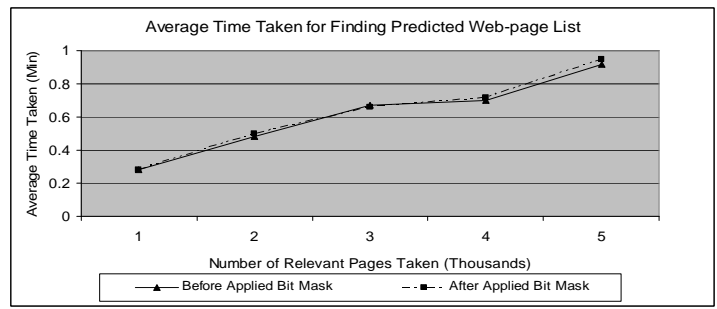

**Fig.4.** Comparison between Average Time Taken for Searching Web-Pages from before and After Bit Masking in IBAG Model

**Accuracy Measure.** To measure accuracy we have used a metric called Harvest Rate *(HR)*. We define HR such as given below:

$$HR := T_{RelSR} / T_{RelSW}$$

Where, $T_{RelSR}$ denotes average of search string term relevance value of all Web-pages exists in search result. $T_{RelSW}$ denotes average of search string term relevance value of all Web-pages selected based on the user given relevance range. While measuring accuracy we have chosen [Maximum Relevance Value, Minimum Relevance Value] as relevance range. Higher value HR denotes more accurate result. In Table 3, we have given an accuracy measure statistics for few search string and observed that accuracy varies on user given search string, but all the cases after bit mask we have achieved better accuracy.

**Table 3.** Accuracy Measure Statistics

| Search String | Number of search result delivered from User Interface | Harvest Rate before bit mask | Harvest Rate after bit mask |
|---|---|---|---|
| ICC player rankings | 20 | 0.296 | 1.467 |
|  | 50 | 0.437 | 1.180 |
|  | 100 | 0.296 | 1.063 |
| best batsman in the world | 20 | 0.590 | 2.128 |
|  | 50 | 0.487 | 1.720 |
|  | 100 | 0.744 | 1.462 |
| ICC world cup 2011 | 20 | 0.358 | 1.490 |
|  | 50 | 0.430 | 1.186 |
|  | 100 | 0.358 | 1.100 |

**Discussion of Average-Case Time Complexity for generating Search Results from both IBAG Model.** To retrieve all the Web-pages in a particular level from IBAG model, we need to traverse $[(1+0) + (1+1) + (1+2) + \ldots + (1+ (n/m - 1))]$ = $[1+2+3+ \ldots + n/m]$ number of Web-pages. We assume that 'n' numbers of Web-pages are distributed in 'm' number of Mean Relevance Level. For finding all Web-pages from IBAG model, we need to traverse $[(1 + 2 + 3 + \ldots + n/m) + (1 + 2 + 3 + \ldots + n/m) + (1 + 2 + 3 + \ldots + n/m) + \ldots$ m times] number of Web-pages. Now, finding a single Web-page from IBAG model in an average case scenario should be:

$$(1/n) \sum_{Level=1}^{m} [1 + 2 + 3 + \ldots + (n/m)] = (1/n) \sum_{Level=1}^{m} [((n/m)*(n/m+ 1))/2]$$
$$= \sum_{Level=1}^{m} [(n/m + 1)/(2*m)] = m*[(n/m + 1)/(2*m)]$$
$$= (n/m + 1)/2 < (n/m) \; \forall n>0, m>0 \text{ and } n>m \approx O(n/m).$$

Say, 'k' number of Web-pages selected from IBAG model for a user given search relevance range. Then the average case time complexity to retrieve 'k' number of Web-pages from IBAG model on which bit masking not applied is $k*O(n/m)$.

The average case time complexity for generating predicted Web-page list from IBAG model on which bit masking applied is given below:

$$p*c* k*O(n/m)$$

Where, k*O(n/m) denotes average case time complexity of finding 'k' number of Web-pages from IBAG model based on the user given search relevance range. 'p' denotes number of Ontology term exist in search string. 'c' denotes time taken for bit operation. Generally, 'p' and 'c' are very very less than 'k' and p*c*k ≈ k. That time the complexity of producing predicted Web-page list becomes k*O(n/m).

## 6   Conclusions

In this paper, we have shown a prototype of multiple Ontology supported Web search engine, which filters search results again and again to present more accurate result to the end users. Basically it retrieves Web-pages from IBAG model. IBAG Web-pages are related to some domains, and our algorithm applied on this specified IBAG is also related to the same domains. Overall, the proposed algorithms have shown the mechanism to generate the bit pattern of all the Web-pages existing in IBAG and as a result prepare a predicted Web-page list using Boolean bit mask.

## References

1. Berners-Lee, T.: "Weaving the Web:The Original Design and Ultimate Destiny of the World Wide Web by its Inventor"; New York, Harper SanFrancisco; 1999
2. Coffman, K. G., Odlyzko, A. M.: "The size and growth rate of the Internet"; AT&T Labs;1998;http://www.dtc.umn.edu/~odlyzko/doc/internet.size.pdf. Retrieved 2007-05-21
3. Khalil, F., Li, J., Wang, H.: "Integrating Recommendation Models for Improved Web Page Prediction Accuracy"; Thirty-First Australasian Computer Science Conference (ACSC); Wollongong; Australia;2008
4. Srivastava, J., Cooley, R., Deshpande, M., Tan, P.: "Web usage mining: Discovery and applications of usage patterns from web data"; SIGDD Explorations 1(2); pp 12-23; 2000
5. Boole, G. [1854]: "An Investigation of the Laws of Thought"; Prometheus Books; ISBN   978-1-59102-089-9; 2003
6. Mukhopadhyay, D., Sinha, S.: A New Approach to Design Graph Based Search Engine for Multiple Domains Using Different Ontologies;11th International Conference on Information Technology, ICIT 2008 Proceedings; Bhubaneswar, India; IEEE Computer Society Press, California, USA; December 17-20, 2008; pp.267-272
7. Gangemi, A., Navigli, R., Velardi, P.: "The OntoWordNet Project: Extension and Axiomatization of Conceptual Relations in WordNet"; In Proc. of International Conference on Ontologies, Databases and Applications of SEmantics (ODBASE 2003), Catania, Sicily (Italy), 2003, pp. 820-838
8. Mukhopadhyay, D., Kundu, A., Sinha, S.: "Introducing Dynamic Ranking on Web-pages based on Multiple Ontology supported Domains"; International Conference on Distributed Computing & Internet Technology, ICDCIT 2010 Proceedings, Bhubaneswar, India; Lecture Notes in Computer Science Series, Springer-Verlag, Germany; February 15-17, 2010; LNCS 5966 pp.104-109